\documentclass[aps,prd,showpacs,superscriptaddress,10pt,twocolumn]{revtex4-2}

\usepackage[utf8]{inputenc}
\usepackage[english]{babel}
\usepackage[T1]{fontenc}
\usepackage{lmodern}

\usepackage{amsmath,amsfonts,amssymb,amsthm,bm,times,dcolumn}

\usepackage{microtype}
\usepackage{braket}
\usepackage{gensymb} 
\usepackage{physics}

\usepackage[colorlinks={true}, citecolor={blue}, filecolor={blue}, 
linkcolor={blue}, urlcolor={blue}]{hyperref}
\usepackage{graphicx,color}
\usepackage[caption=false]{subfig}

\usepackage{soul}
\usepackage[normalem]{ulem}

\begin{document}
	

\title{Nonclassical features of the pointer states in the $q$-deformed post-selected weak measurement}

\author{Seyyede Elham Mousavigharalari}\email{elham.mousavi@istanbul.edu.tr}
\affiliation{Department of Physics, Istanbul University, Vezneciler, Istanbul 34134, T\"urkiye}
\author{Azmi Ali Altintas}
\affiliation{Department of Physics, Istanbul University, Vezneciler, Istanbul 34134, T\"urkiye}

\author{Fatih Ozaydin}
\affiliation{Institute for International Strategy, Tokyo International University, 4-42-31 Higashi-Ikebukuro, Toshima-ku, Tokyo 170-0013, Japan}

\date{\today}

\begin{abstract}
We study $q$-deformed coherent states of the Arik-Coon harmonic oscillator as the quantum resource in the post-selected weak measurement. 
First, we show how the precision of weak measurement is improved significantly due to $q$-deformation. 
Next, we focus on the role of the interplay between the deformation parameter and the interaction strength on the nonclassical nature of light.
In particular, we show that sub-Poissonian photon distribution as characterized by Mandel parameter,  photon antibunching effect, and quadrature squeezing are greatly enhanced by $q$-deformation.
Our results not only advance the understanding of the fundamentals of $q$-deformed quantum mechanics, but also raise the potential to contribute to quantum technologies.
\end{abstract}

\maketitle

\section{Introduction}\label{sec:introduction}
In recent decades, q-deformation has garnered escalating interest across multiple domains of physics, including but not limited to quantum mechanics and quantum optics~\cite{altintas2009inhomogeneous,altintas2010inhomogeneous,altintas2011inhomogeneous,chaichian1990quantum,arikan2024one,sargolzaeipor2019q,bayindir2021self,bayindir2022petviashvili}, quantum information~\cite{altintas2014constructing,altintas2020q,samar2023modelling}, nuclear physics~\cite{bonatsos1999quantum}, quantum thermodynamics~\cite{lavagno2002generalized,tuncer2019work,gupta1994planck,ozaydin2023powering,naseri2022non}, and statistical mechanics~\cite{algin2017effective,boumali2023thermal}. 
This stems from the fact that it provides a richer structure for exhibiting nonclassical features and also makes it possible to obtain a deeper understanding of the physical systems by giving the system an additional degree of freedom~\cite{meng2007q,berrada2012bipartite,dodonov2002nonclassical,chung2019q}.
Considering the widespread application of quantum optical states in almost every branch of physics, the study of the generalization of these states has become one of the most important tasks in physics. 

Initial efforts were concentrated on generalization of the coherent states within the framework of group theory~\cite{perelomov1972coherent}. 
While mathematical challenges have impeded the pace of this progression, significant achievements have been made. The 
$q$-generalization offers a systematic framework that establishes a valid quantum optics approach for diverse deformed quantum mechanical models.
The initial generalization of the simple harmonic oscillator using $q$-boson annihilation and creation operators was introduced by Arik and Coon in 1976, with $q\rightarrow1$ corresponding to the non-deformed scenario~\cite{arik1976hilbert}. 

Subsequently, numerous authors have proposed diverse forms of $q$-deformations for the simple harmonic oscillator and investigated their associated novel quantum states, including $q$-coherent states, $q$-cat states, and $q$-squeezed states. 
It has been observed that these states exhibit distinct behaviors, commonly referred to as quantum or nonclassical effects, which are contingent upon the deformation parameter $q$~\cite{buvzek1992jaynes, damaskinsky1991deformed,kowalski1993coherent, mcdermott1994analogue, damaskinskii1995hermite, perelomov1996completeness, kwek1998general,parthasarathy2002diagonal,eremin2006q, ching2013generalized, gazeau2013pisot,dey2013time, dey2014noncommutative, dey2015q, zelaya2018generalized, dey2020generalized}. 
According to the conventional quantum theory, a quantum measurement, also known as the strong measurement, is a set of probabilistic orthogonal projections onto the eigenstates of an observable. 
In such a measurement, first a physical property of a system we intend to measure, i.e., the observable $\hat{A}$ is coupled to the meter, and then the meter is used to read out information about the state of the observable. 

Since the strong measurement is not time-symmetric, in $1988$ Aharonov, Albert and Vaidman (AAV) proposed a different quantum measurement technique called the post-selected weak measurement (PSWM)~\cite{aharonov1988result}. 
It implies that when considering time-symmetric measurement, it becomes necessary to post-select the measuring system subsequent to the measurement interaction~\cite{aharonov1964time}. 
The primary assumption underlying their discussion is that the coupling between the meter's probe (the pointer) and the physical property of the system that is being measured (the observable) is sufficiently weak that the entanglement between them, which typically arises in standard measurements, can be disregarded. 
In standard quantum measurement, the output of a quantum measurement or the shift of the meter is typically within the range of the minimal and the maximal eigenvalues of $\hat{A}$. 
Nonetheless, in weak measurement, the meter shift is smaller than the uncertainty associated with the observables of the measurement apparatus, rendering it challenging to interpret. 
Consequently, achieving a precise estimation of the desired parameter becomes unfeasible. 

However, when appropriate pre- and post-selections are applied to the state of the observable $\hat{A}$ following the weak coupling, the measurement outcome is enhanced by the post-selection. 
The magnitude of this enhancement can be quantified by a parameter known as the ``weak value'', which significantly surpasses the range of eigenvalues associated with the observable $\hat{A}$.
In strong measurements, the outcome might be random, while for the weak measurements it has always a definite value and is equal to the real part of the weak value~\cite{pang2015improving,pang2016protecting,kedem2012using,starling2009optimizing,nishizawa2012weak,jordan2014technical,chen2021beating}.   
An important aspect of weak measurement is its relation to nonclassicality witnesses and advantages in quantum metrology~\cite{arvidsson2021conditions}, where surpassing the classical limit via nonclassical resources is crucial~\cite{pezze2009entanglement,erol2014analysis,ozaydin2015quantum,ozaydin2020parameter}.
Weak measurement has also enabled direct observation of Hardy's paradox~\cite{yokota2009direct}.

Nonclassical light is a potential resource to overcome classical limitations in high-precision measurements~\cite{tan2019nonclassical}. 
For instance, the use of squeezed light which represents prominent nonclassical properties, has shown to enhance measurement sensitivities beyond the classical limits. This has emerged as a fundamental concept in quantum metrology~\cite{li2015weak,pezze2018quantum,gessner2020multiparameter,araya2021influence}. 
Considering post-selection and nonclassicality of the probe states as two main resources to achieve high precision in a weak measurement regime, we will exploit the advantage of more nonclassical $q$-deformed quantum resources as the probe states~\cite{turek2015post,turek2020effects,aishan2022amplitude,hu2022quantum,xu2022studying,turek2023single}. 
Numerous authors have shown that by tuning the deformation parameter $q$, the $q$-deformed coherent states (in brief, $q$-coherent states) will offer a richer structure, exhibiting nonclassical properties such as sub-Poissonian photon distribution and antibunching and quadrature squeezing in comparison to their non-deformed counterparts~\cite{quesne2002new,quesne2003maths,quesne2003geometrical,fakhri2016nonclassical,fakhri2016symmetric,fakhri2020nonclassical,fakhri2021q}. 

In this work, the non-orthogonal $q$-coherent states  of the Arik-Coon harmonic oscillator will be utilized as the probe states in the post-selected weak measurement.
This paper is organized as follows. 
In Sec.\ref{sec:Generalization} we will go beyond the state of the art and will study the weak values of $q$-deformed position operator in the PSWM by making use of the $q$-coherent states of the Arik-Coon harmonic oscillator as the pre- and post-selected states of the pointer to determine the role of $q$-deformation on the weak value amplification effect. 
The goal is pursued by investigating the nonclassical properties of the $q$-coherent state pointers including sub-Poissonian photon statistics, anti-bunching effect, and amplitude and quadrature squeezing in the PSWM in Sec.~\ref{sec:Nonclassical}. 
Finally, we conclude in Sec.~\ref{sec:Conclusion}.

\section{$q$-generalized post-selection}\label{sec:Generalization}
$q$-deformed nonlinear coherent states were first introduced by Arik-Coon as the eigenstates of the $q$-deformed annihilation operator, i.e., ${a_{q}}|z\rangle_{_{q}}=z|z\rangle_{_{q}}$, where $z$ is a complex variable with the polar form $z=|z| e^{i \theta}$.
Post-selected weak measurement basically involves three steps: 

\textit{1)} Pre-selection, i.e., the preparation of the measuring system and the measuring probe in specific initial states 
$|\psi_{i}\rangle$ 
and 
$|\phi\rangle$, 
respectively. 

\textit{2)} Weak interaction between the system and the probe through a unitary transformation which will lead to the updated state $|\Psi\rangle=\hat{U}|\psi_{i}\rangle$ 
with the unitary operator 
$\hat{U}=e^{{-i \int_{t_0}^{t} H_{\text{int}} dt}}$. 
Here, 
$H_{\text{int}}=g(t) \hat{A}\otimes \hat{P}$ 
is the interaction Hamiltonian between the system that is being measured and the meter probe (pointer), where $\hat{A}$ is the observable of the system to be measured and $\hat{P}$ is the momentum operator relevant to the pointer. 
Moreover, the coupling 
$g(t)=g \delta(t-t_0)$ 
is a nonzero function in a finite interaction time interval $t$ with $g$, a small real coupling constant, and $\delta(t-t_0)$ implies that the weak interaction occurs over a very brief duration of time. 

\textit{3)} Post-selection, i.e., performing a strong measurement and observing the system such that if it is found in the specific preferred state $|\psi_{f}\rangle$ of the probe’s pointer, it will be recorded, and otherwise, it is ruled out and in this way the final state of the pointer, $|\Phi\rangle$ will be obtained~\cite{aharonov1964time}.

The main result of AAV’s measurement regime is that for a real coupling constant $g$, during the interaction time $t$, the weak interaction strength $gt$ (note that it is different from $g(t)$) is sufficiently weak after the post-selection. Therefore, the mean shift of the pointer position is given by $\Delta {x}=g t \text{Re} (A_w)$, with $\text{Re}$ denoting the real part, and $A_w$ is the so-called weak value 
\begin{equation}\label{Weakvalue Def.}
	{A_w}=\frac{\langle \psi_f |\hat{A} |\psi_i\rangle}{\langle \psi_f | \psi_i \rangle }.
\end{equation} 

In other words, weak value is the average shift of the final pointer state or the information obtained about the system between the initial and final states.
 
In particular, the weak value can be complex with the real and imaginary parts that contain distinct information about the system~\cite{steinberg1995conditional}. 
The real part represents the value measured and registered by the pointer~\cite{johansen2004nonclassicality}, while the imaginary part corresponds to the displacement in the momentum-space distribution of the meter indicating that how the initial state of the system would be perturbed unitarily~\cite{gedik2021weak}. 
It was found in a recent study by Guo et al. that the pointer states with nonclassical features result in a larger average pointer shifts and a wider range of weak values~\cite{guo2021enhancing}, so, the authors believe that this leads to higher accuracy compared with using ordinary state as the pointer state. 

$q$-deformed quantum field theory, which violates the Pauli exclusion principle and deviates from Bose statistics~\cite{greenberg1989phenomenology,mohapatra1990infinite}, has led to the formulation of $q$-deformed harmonic oscillator algebra by introducing the deformation parameter $q=e^{\hbar}$ into the canonical commutation relations. These $q$-deformed simple harmonic oscillators, often referred to as $q$-oscillators, have garnered significant attention in physics due to two distinct properties: their significant impact in completely integrable theories and their connection with a specific form of nonlinearity arising from the dependence of the oscillation frequency on the energy of the vibrations~\cite{dey2015q,meng2007q,zelaya2018generalized,fakhri2020nonclassical, fakhri2021q}.

The two primary commutation relations $a a^{\dagger} -qa^{\dagger} a=1 $ and  $a a^{\dagger} -qa^{\dagger} a=q^{-N} $, with $0<q<1$ are associated with the Math-type ($M$-type) and Physics-type ($P$-type) $q$-deformed quantum harmonic oscillators in the literature introduced by Arik and Coon~\cite{arik1976hilbert}, and Biedenharn~\cite{biedenharn1989quantum} and Macfarlane~\cite{macfarlane1989q}, respectively. 
According to Katriel and Solomon's model~\cite{katriel1994nonideal}, $q$-coherent states provide a more accurate description of non-ideal $\text{He-Ne}$ lasers, and it is hypothesized that the deformation parameter $q$ functions as a tuning parameter that defines the degree of deviation of the implemented device from an ideal one.

$q$-deformed systems exhibit specific characteristics wherein they manifest more nonclassical properties in comparison to their non-deformed counterparts. Furthermore, an intriguing aspect is that the extent of nonclassicality is heavily contingent upon the value of the deformation parameter. Put differently, the deformation parameter $q$ acts akin to an additional degree of freedom for the system, thereby enhancing its nonclassical properties.

The exact form of the q-coherent state is 
\begin{equation}
	|z\rangle_{q}=\frac{1}{\sqrt{e_{q} {(|z|^2)}}} \sum_{n=0}^{\infty} \frac{z^n}{\sqrt{{[n]_q}!}} |n\rangle,
\end{equation}
\noindent
with ${[n]_q}=\frac{1-q^n}{1-q}$, the relative $q$-number of the M-type bosons~\cite{fakhri2016nonclassical}.
The $q$-generalized exponential function is of the form  
\begin{equation}
	e_{q} ({|z|}^2)=\sum_{n=0}^{\infty} \frac{{|z|^{2n}}}{{[n]_q}!}=\sum_{n=0}^{\infty} \frac{((1-q)\, |z|^2)^n}{(q;q)_n},
\end{equation}
\noindent
where ${(a;q)_n:=(1-a)(1-aq)(1-aq^2)\cdots(1-aq^{n-1})}$ and ${(a;q)_0}=1$. 
Note that any result regarding the $q$-deformed coherent states of the probes are required to reduce to their non-deformed counterparts in ordinary quantum mechanics, by taking the deformation parameter $q\rightarrow1$.

Considering the $q$-coherent states of the Arik-Coon harmonic oscillator in PSWM as the probe is expected to bring about more nonclassicality through giving the system two extra degrees of freedom, one being the deformation parameter $q$ and the other, the coupling constant $g$, both playing a key role in improving the nonclassicality of the system which is necessary to obtain more precision in quantum measurement.

Let us introduce the $q$-generalized post-selected weak measurement considering the $q$-coherent states as the initial and final states of the system and pointer. 
The probe will be coupled to the system with an interaction Hamiltonian of the form 
\begin{equation}
	H_{\text{int},q}=g(t) \hat{A_{q}}\otimes \hat{P_{q}},
\end{equation}
\noindent
where $\hat{A_{q}}$ and $\hat{P_{q}}$ are the $q$-deformed position and momentum quadratures of the system and probe, respectively. 
Hence, the unitary evolution operator in terms of the $q$-exponential function turns out to be
\begin{equation}
	\begin{split}
	\hat{U}_{q} & = e_{q}^{-i \int_{t_0}^{t} H_{\text{int.,}q} \, dt} \\ 
	& = \sum_{n=0}^{\infty} \frac{{\left( (1-q)(-ig\hat{A}_{q}\otimes\hat{P}_{q}) \right) ^n}}{{(q;q)_{n}}}.
	\end{split}
\end{equation}
Suppose that the system and probe are initially prepared in the states $|\psi_{i}\rangle_{q}$ and $|\phi\rangle_{q}$, respectively.
Then the overall state of the joint system after interaction will be turn into
\begin{equation}
	|\Psi\rangle_{q}=\hat{U_{q}} |\psi_{i}\rangle_{q} {|\phi\rangle}_q.
\end{equation}
When the interaction is sufficiently weak, only the lowest order terms of $gt$ are considered. 
Then $\hat{U}_{q}$ is obtained as

\begin{equation}
	\hat{U_{q}}=1-i g \hat{A_{q}}\otimes \hat{P_{q}}.
\end{equation}

\noindent
Post-selection of the system to the final state ${|\psi\rangle}_{f}$ leads to
\begin{equation}
		{|\Phi\rangle}_{q}= {_{q}\langle}{ \psi_{_f}}|\Psi\rangle_{q},
\end{equation}

\noindent
and therefore the final state of the probe will be

\begin{equation}
		|\Phi\rangle_{_{q}}={\cal{N}}\, \,_{_{q}}\langle \psi_{f}|\psi_{i}\rangle_{_{q}}\left( |\phi\rangle_{_{q}} -i g \langle \hat{A}_{q}\rangle _{_{w}}\otimes \hat{P}_{q} |\phi\rangle_{_{q}} \right),	
\end{equation}

\noindent
with 
$\langle \hat{A}_{q}\rangle _{_{w}}=\frac{_{_{q}}\langle \psi_{f}|{\hat{A}}_q|\psi_{i}\rangle_{_{q}}}{_{_{q}}\langle \psi_{f}|\psi_{i}\rangle_{_{q}}}$
being the weak value of the observable ${\hat{A}}_q$ to be measured.
The normalization factor for $|\Phi\rangle_{f}$ can be found as 
\ \\ 

\begin{widetext}
	\begin{equation}
		{\cal{N}}(g,q)  =\left[ {|_{_{q}}\langle \psi_{f}|\psi_{i}\rangle_{_{q}}|}^2 \ \left( 1 
		 +2\sqrt{2} \  \text{Im}(z) \, \text{Im}(g \langle \hat{A}_{q}\rangle_{_{w}})
			 + \frac{g^2}{2} {|\langle \hat{A}_{q}\rangle _{_{w}}|}^2\, 
			 \left( 1+(1+q) {|z|}^2-2 {({\mbox{Re}}(z))}^2 \,+2 {(\mbox{Im}(z))}^2 \right)     \right) \right]^{-\frac{1}{2}}.
\end{equation}
\end{widetext}

In this work we consider measuring two $q$-generalized position operators  
$ \hat{A}_{_{1,q}}= \hat{X}_{_{1,q}}=\frac{1}{\sqrt{2}} ({{a_q}^\dagger}+{a_q})$ 
and 
$\hat{A}_{_{2,q}}= \hat{X}_{_{2,q}}=\frac{1}{\sqrt{2}} ({q^{\frac{N}{2}}} \, {a_q}^\dagger + \,  {a_q} \, q^{\frac{N}{2}}  )$ 
via a  $q$-deformed pointer with momentum quadrature $\hat{P_q}=\frac{i}{\sqrt{2}} ({{\hat{a}}_{q}}^\dagger - {\hat{a}}_q)$. 

Let us suppose the initial and final states of the system are $|\psi_{i}\rangle_{_{q}}=|\alpha\rangle_{_{q}}$ and $|\psi_{f}\rangle_{_{q}}=|\beta\rangle_{_{q}}$, respectively. Then, it is straightforward to calculate the weak values of  $\hat{X}_{_{1,q}}$ and $\hat{X}_{_{2,q}}$ from Eq.~\ref{Weakvalue Def.}, as

\begin{equation}
	\langle \hat{X}_{_{1,q}}\rangle_{_{w}}=\frac{ (\alpha + {\beta}^*)}{\sqrt{2}},
\end{equation}

\noindent
and

\begin{equation}
	 \langle \hat{X}_{_{2,q}}\rangle_{_{w}}={ \frac{{q^\frac{1}{2}} \, {e_q}({q^\frac{1}{2}} \alpha {\beta}^*)}{{e_q}(\alpha {\beta}^*)}}{\frac{(\alpha + {\beta}^*)}{\sqrt{2}}}
\end{equation}

\noindent
with $\alpha=|\alpha|e^{i\lambda}$ and $\beta=|\beta| e^{i\varphi}$.
As it is seen, the weak values of both observables depend on the coherency variables $|\alpha|$ and ${|\beta|}$, and the angles $\lambda$ and $\varphi$. The most important point is that $\langle \hat{X}_{_{1,q}}\rangle_{_{w}}$ is $q$-independent while $ \langle \hat{X}_{_{2,q}}\rangle_{_{w}}$ is completely $q$-dependent.
It should be remarked that the fidelity between the initial and final states of the system used in the above calculation is

\begin{equation}
	_{_{q}}\langle \beta|\alpha\rangle_{_{q}}=\frac{{e_{q}} ({\beta}^* \alpha )}{\sqrt{{e_q}(|\beta|^2) {e_q}(|\alpha|^2)}}.
\end{equation}

In ordinary systems, the desired fidelity between the pre-and post-selection is a small one, so the weak value can be very large, indicating a large shift in the pointer position which can be a powerful tool to amplify the weak signals and detecting the weak effects~\cite{hosten2008observation}. 
Here, the dependency of the fidelity to the deformation parameter gives the system the opportunity for obtaining large weak values by tuning the value of $q$. 

\begin{figure}[t!]
	\centering
	\includegraphics[width=1\linewidth]{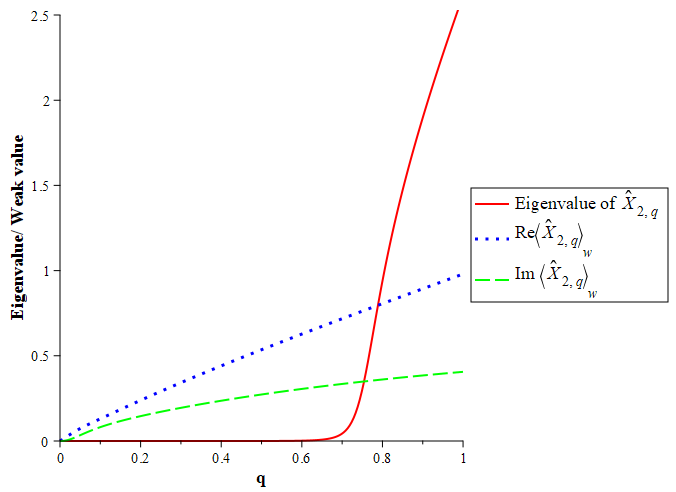}
	\caption{Eigenvalue and complex and real parts of the complex weak values of the observable $\langle\hat{X}_{_{2,q}}\rangle_{_{w}}$ with respect to deformation parameter $q$ with $|\lambda|=2$, $|\varphi|=0.5$, $\theta=\frac{\pi}{8}$, $\lambda=\frac{7 \pi}{8}$.}
	\label{fig:Fig1}
\end{figure}

Fig.~\ref{fig:Fig1} demonstrates the amplification of the $q$-dependent complex weak value with respect to the deformation parameter $q$, showing that both the real and imaginary parts of the weak value are greater than the eigenvalues for $q  \lessapprox 0.8$, and the weak value amplification due to $q$-deformation is clearly confirmed. 
For the range $0.8 \lessapprox q<1$, this trend is reversed and the eigenvalues are greater than the weak values.

Therefore, based on experimental and theoretical evidences~\cite{guo2021enhancing,kedem2012using}, this system can effectively suppress technical noise by amplifying the signal by tuning the deformation parameter properly. 
This consequently leads to more precise measurements and enriches the ability to detect weak physical effects.

\section{Nonclassical properties of the $q$-coherent state pointers}\label{sec:Nonclassical}
In this section we will focus on the effects of weak measurement on the nonclassical features of the coherent states of the $q$-deformed radiation fields. 
We will first study the photon statistical properties of these states, namely the super-(sub-)Poissonian statistics. 
Deviation of the standard photon statistics from Poisson distribution is characterized by the Mandel parameter, $Q$, such that $Q=0$ corresponds to Poissonian statistics, whereas $Q<0$ and $Q>0$ indicates sub- and super- Poissonian statistics, respectively~\cite{kryuchkyan2003sub}. 
Sub-Poissonian radiation is called nonclassical light, since the number of photons distribution on a state is narrower than a Poissonian one with the same intensity~\cite{davidovich1996sub}.
We will first show how the nonclassical nature of light can be enhanced with $q$-deformation.

Next, we will explore photon antibunching  as a pure quantum effect~\cite{paul1982photon,koashi1993photon}. 
Second-order correlation function which represents the probability of two photons arriving simultaneously from a random source can be used to investigate the photon antibunching, a coherent state of a laser beam and bunching effect of a classical field. 
Observation of antibunching trait reveals whether there is only a single photon emitter present in a sample. 
We will show that $q$-deformation leads to stronger photon antibunching.

The last nonclassical characteristics we will study is the quadrature squeezing corresponding to the case where the variance of at least one of the quadratures $X$ or $P$ is less than the right hand side of the Weyl-Heisenberg uncertainty relation, $\sigma_{XX}\,\sigma_{PP}\geq \frac{1}{4} {|[\hat{X}_{q},\hat{P}_{q}]|}^2$~\cite{buvzek1992superpositions}.  
According to the uncertainty relation, when there is squeezing in one of the quadratures, the other quadrature is expanded. 
Here, we will show how the quadrature squeezing is enhanced due to $q$-deformation.

\begin{figure}[t!]
	\begin{center}
		\includegraphics[width=1\linewidth]{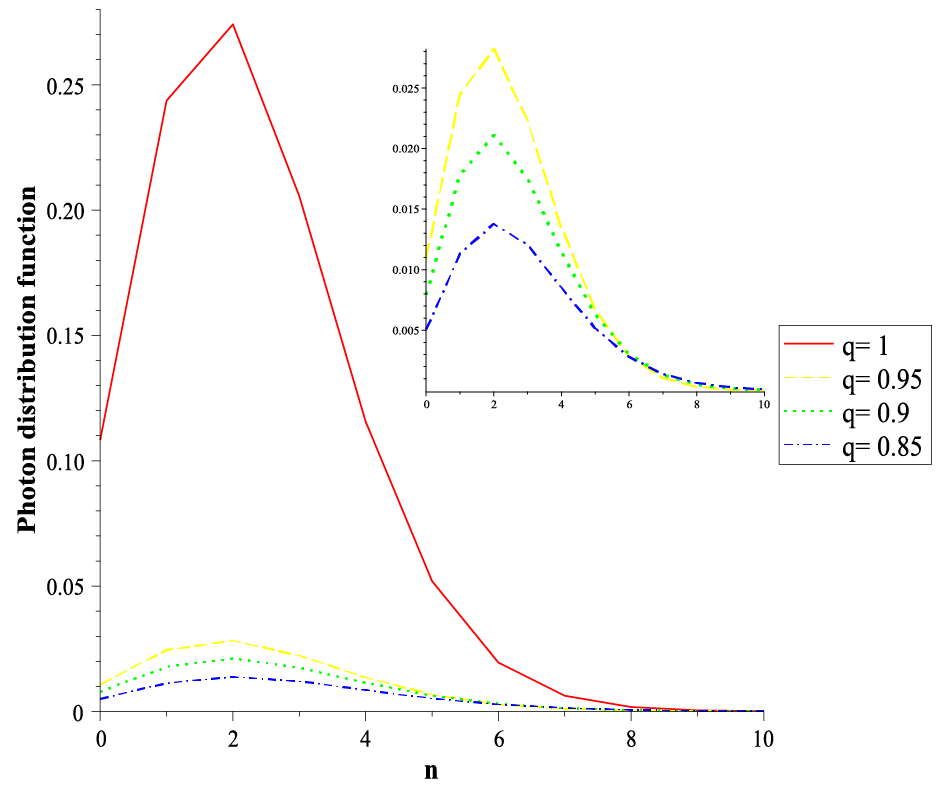}
	\end{center}
	\caption{Photon distribution function for the $q$-coherent state probe for the weak value $\langle\hat{X}_{_{2,q}}\rangle_{_{w}}$ with various instances of $q$, and  $g=0.3$, $|z|=1.5$, $|\alpha|=2$, $|\beta|=1$,  $\theta=\lambda=\varphi=\frac{\pi}{2}$.} \label{fig:Fig2}
\end{figure}

\subsection{Sub-Poissonian behavior and photon number squeezing}
For an arbitrary non-negative integer $n$, the distribution function or the conditional probability of detecting $n$ photons in a weak measurement post-selected with a q-deformed coherent states probe ${|\Phi \rangle}_{_{q}}$ can be found for $\langle\hat{X}_{_{1,q}}\rangle_{_{w}}$ and $\langle\hat{X}_{_{2,q}}\rangle_{_{w}}$ as

\begin{equation}
	\begin{split}
		{P_{{\langle\hat{X}_{_{1,q}}\rangle}_{{w}}}}(q;n) &\equiv \left|_{_{q}}\langle n|\Phi\rangle _{_{q}}\right|_{w_1}^2 \\
		&= {{|\cal N|}^{2}} \frac{{e_{_{q}}(\alpha \beta^*)} {e_{_{q}}(\alpha^* \beta)}}{e_{_{q}}(|\alpha|^2) e_{_{q}}(|\beta|^2) e_{_{q}}(|z|^2)} \frac{|z|^{2n}}{{[n]_{q}}!} \\
		& \cross \left(1 - g \mbox{Re}((\alpha + \beta^*) z) + \frac{g^2}{4} |\alpha + \beta^*|^2 |z|^2\right) ,
	\end{split}
\end{equation}
\noindent
and 
\begin{equation}
	\begin{split}
		{P_{{\langle\hat{X}_{_{2,q}}\rangle}_{{w}}}}(q;n) &\equiv \left|_{_{q}}\langle n|\Phi\rangle _{_{q}}\right|_{w_2}^2 \\
		&= {{|\cal N|}^{2}} \frac{{e_{_{q}}(\alpha \beta^*)} {e_{_{q}}(\alpha^* \beta)}}{e_{_{q}}(|\alpha|^2) e_{_{q}}(|\beta|^2) e_{_{q}}(|z|^2)} \frac{|z|^{2n}}{{[n]_{q}}!} \\
		& \Bigg(1 - g \mathrm{Re}\left(\frac{{q^\frac{1}{2}}\,{e_q{({q^\frac{1}{2}}\alpha \beta^*})}}{e_q{(\alpha \beta^*)}} {(\alpha + \beta^*) z}\right) \\
		&+ \frac{g^2}{4} {|\alpha + \beta^*|^2  \frac{q\, e_q{({q^\frac{1}{2}}\alpha \beta^*)}\,e_q{({q^\frac{1}{2}}\alpha^* \beta})}{e_q{(\alpha \beta^*)}\, e_q{(\alpha^* \beta)}}}|z|^2 \bigg).
	\end{split}
\end{equation}
\vfill\pagebreak
\onecolumngrid
\onecolumngrid
\begin{figure}[t!]
	\begin{center}
		\subfloat[]{\includegraphics[width=0.4\linewidth]{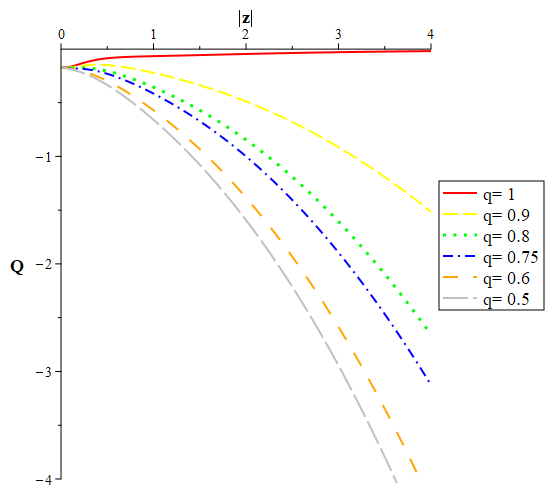}} \
		\subfloat[]{\includegraphics[width=0.4\linewidth]{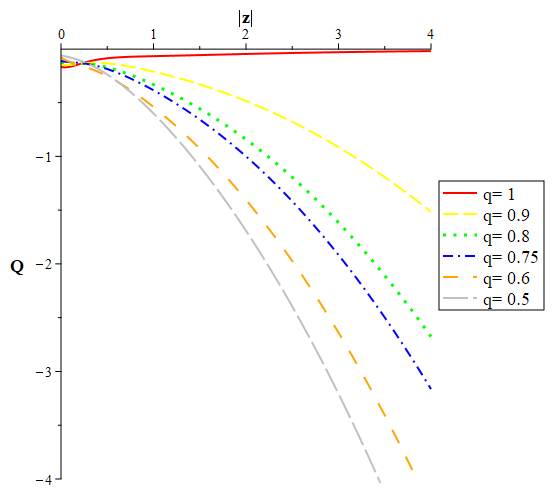}} \\
		\subfloat[]{\includegraphics[width=0.4\linewidth]{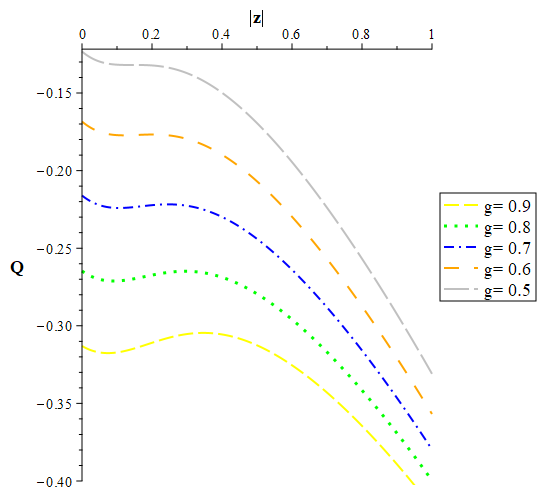}} \
		\subfloat[]{\includegraphics[width=0.4\linewidth]{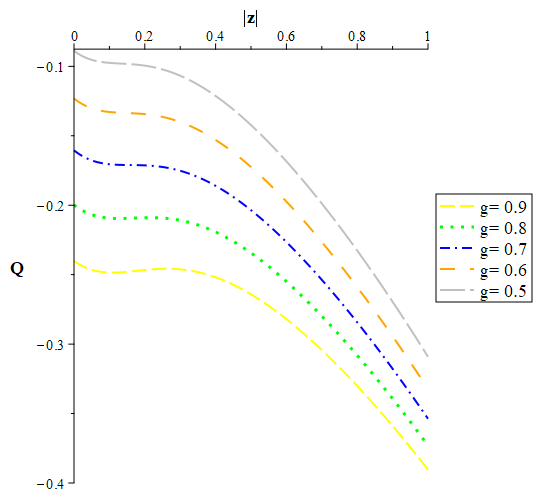}}
	\end{center}
	\caption{$q$-Mandel parameter for the $q$-coherent state probes with (a) $\langle\hat{X}_{_{1,q}}\rangle_{_{w}}$ with $g=0.6$, (b) $\langle\hat{X}_{_{2,q}}\rangle_{_{w}}$ with $g=0.6$, (c) $\langle\hat{X}_{_{1,q}}\rangle_{_{w}}$ with $q=0.8$, and (d) $\langle\hat{X}_{_{2,q}}\rangle_{_{w}}$ with $q=0.8$, and all with $|\alpha|=2$, $|\beta|=0.5$ and $\theta=\frac{\pi}{2}$, $\lambda=\frac{\pi}{8}$, $\varphi=\frac{7\pi}{8}$.}\label{fig:Fig3}
\end{figure}
\twocolumngrid

Variations of the above photon distribution function ${P_{{\langle\hat{X}_{_{2,q}}\rangle}_{{w}}}}(q;n)$ are presented in Fig.~\ref{fig:Fig2}, while the former function is omitted due to the strong similarity of the two plots.
The height of the $q$-post-selected photon distribution function 
decreases with decreasing $q$. 
Moreover, the peaks slightly shift to the left. 
Deviation of the distribution function from the Poissonian reveals that the deformation parameter $q$ affects the distribution significantly, enhancing the nonclassical features.
Mandel parameter $Q$ is an appropriate measure of deviation from the Poisson distribution that identifies the nonclassical nature of photon probability distribution function~\cite{kryuchkyan2003sub,mandel1995optical,gerry2023introductory}. 
Hence, we consider $Q$ in the $q$-deformed setting as a $q$-Mandel parameter as in~\cite{fakhri2020nonclassical,fakhri2021q}
\begin{equation}\label{mandel}
	\begin{split}
	Q_{{{|\Phi \rangle}_{q}}}^{({\cal A})} & \equiv \frac{\langle (\Delta \hat{n})^2\rangle }{\langle \hat{n}\rangle} -1 \\
	& = \frac{{_{_{q}}\langle \Phi|}({{\cal A}^\dagger}{\cal A})^2{{|\Phi \rangle _{_{q}}}}-{ {_{_{q}}\langle \Phi|}{{\cal A}^\dagger}{\cal A}{{|\Phi \rangle _{_{q}}}}}^2 }{{_{_{q}}\langle \Phi|}{{\cal A}^\dagger}{\cal A}{{|\Phi \rangle _{_{q}}}}}-1,
	\end{split}
\end{equation}
\noindent
and derive the $q$-Mandel parameter, $Q$ under weak measurement with $q$-deformed coherent state probes. 
Skipping the intermediary steps, we find $Q$ for a weak value $\langle\hat{A}_{_{q}}\rangle_{_{w}}$ as 
\begin{widetext}
\begin{equation}
	\begin{split}
		Q = &\Bigg\{\bigg\{\bigg[ {{|\cal N|}^{2} |_{_{q}}\langle \beta|\alpha\rangle_{_{q}}|^2}  \bigg( {|z|^2}+q|z|^4 +\frac{2g}{\sqrt{2}} \mathrm{\mbox{Re}}\bigg( \langle\hat{A}_{_{q}}\rangle_{_{w}} \left({z^*}+(2 q z^* +q^2 z^* -z ) {|z|^2}+(q^3 z^* -q z) |z|^4\right) \bigg) \\
		&+\frac{g^2}{2} |\langle\hat{A}_{_{q}}\rangle_{_{w}} | ^2\bigg( 1-z^2 -{z^*}^2 -\left((2 q+q^2) {(z^2 +{z^*}^2)} -(q^3+3 q^2+3 q) \right) |z|^2 \\
		&-\left( q^3 ({z^2}+{z^*}^2)-(q^5+2 q^{4}+3 q^{3})-1 \right){|z|^4}+({q^6}+q) {|z|^6}\bigg)\bigg)\bigg] \\
		&-\bigg[ {|\cal N|}^{2} |_{_{q}}\langle \beta|\alpha\rangle_{_{q}}|^2 \bigg( {|z|^2}+\frac{2g}{\sqrt{2}} \mathrm{\mbox{Re}}\left(\langle\hat{A}_{_{q}}\rangle_{_{w}} \left( {z^*}+ (q z^* -z) {|z|^2} \right) \right) \\
		&+\frac{g^2}{2} |\langle\hat{A}_{_{q}}\rangle_{_{w}} |^2 \left(1+(q+2- z^2-{{z^*}^2}){q |z|^2} +(1+q^3) {|z|^4} -{z^2}-{{z^*}^2}\right) \bigg) \bigg] \Bigg\} \\
		&\times \bigg\{\bigg[ {|\cal N|}^{2} |_{_{q}}\langle \beta|\alpha\rangle_{_{q}}|^2 \bigg( {|z|^2}+\frac{2g}{\sqrt{2}} \mathrm{\mbox{Re}}\left(\langle\hat{A}_{_{q}}\rangle_{_{w}} \left( {z^*}+ (q z^* -z) {|z|^2} \right) \right) \\
		&+\frac{g^2}{2} |\langle\hat{A}_{_{q}}\rangle_{_{w}} |^2 \left(1+(q+2 - z^2-{{z^*}^2}){q|z|^2} +(1+q^3) {|z|^4} -{z^2}-{{z^*}^2}\right) \bigg) \bigg]^{-1} \Bigg\} -1. 
	\end{split}
\end{equation}
\end{widetext}

In Figs.~\ref{fig:Fig3}(a) and (b), we present the $q$-Mandel parameter of the $q$-post-selected coherent state probes for $\langle\hat{A}_{_{q}}\rangle_{_{w}}=\langle\hat{X}_{_{1,q}}\rangle_{_{w}}$
and 
$\langle\hat{A}_{_{q}}\rangle_{_{w}}=\langle\hat{X}_{_{2,q}}\rangle_{_{w}}$, respectively,  for a range of $q$ with $g=0.6$. 
In both cases, the distribution is completely sub-Poissonian and therefore it is nonclassical. 
Moreover, the sub-Poissonian character becomes stronger, showing that the nonclassicality is enhanced with decreasing $q$. 

Mandel parameter for $q=0.8$  and a range of interaction parameters $g$ are presented in Figs.~\ref{fig:Fig3}(c) and (d) for $\langle\hat{A}_{_{q}}\rangle_{_{w}}=\langle\hat{X}_{_{1,q}}\rangle_{_{w}}$ and  $\langle\hat{A}_{_{q}}\rangle_{_{w}}=\langle\hat{X}_{_{2,q}}\rangle_{_{w}}$, respectively. 
It can be seen easily by comparing for example the red solid curve in Fig.~\ref{fig:Fig3}(a) corresponding to $q=1$ and $g=0.6$, with the dashed green curve in Fig.~\ref{fig:Fig3}(c) corresponding to $q=0.8$ and $g=0.6$, that for $q<1$, the role of the interaction strength in making the distribution sub-Poissonian becomes more prominent.

Another important point is that a special type of squeezing, the so called photon number (amplitude) squeezing, exists for the $q$-post-selected pointers. 
Photon number squeezing is a behavior exhibited by a sub-Poissonian photon distribution, where the photon number dispersion is narrower than the average number of photons in a state, i.e., $\langle (\Delta \hat{n})^2\rangle < \langle \hat{n}\rangle$~\cite{gerry2023introductory}. 

This property results in optical noise that can be much lower than the classical shot-noise limit, making sub-Poissonian light highly useful for quantum information processing~\cite{dey2014noncommutative}.
Therefore, the advantage of choosing a $q$-deformed space is evident in this context.
In response to stronger interaction and smaller $q$, the magnitude of the negative Mandel parameter becomes greater.
This phenomenon is unique to the $q$-deformed states and is not observed in the ordinary (non-deformed) case presented by the solid red lines in Figs.~\ref{fig:Fig3} (a) and (b).

\subsection{Photon antibunching effect}\label{Antibunching}
The existence of antibunched light is of interest to theoretical physics due to its compatibility with the quantum theory of light and its variance with classical theory~\cite{teich1988photon}.
Second-order correlation function in zero delay time $g^{(2)}(0)$ associated with normalized $q$-coherent state pointers in weak measurement is defined as 

\begin{equation}
	g^{(2)}_{{{|\Phi \rangle}_{q}}}(0)\equiv\frac{{_{q}\langle \Phi|}{{{\cal A}^\dagger}}^2{{\cal A}}^2{{{{|\Phi \rangle}_{q}}}}}{{_{q}\langle \Phi|}{{\cal A}^\dagger}{{\cal A}}{{{{|\Phi \rangle}_{q}}}}^2}.
\end{equation}

\noindent
For the $q$-post-selected pointer state with every weak value
${\langle\hat{A}_{_{q}}\rangle_{_{w}}}$, we find

\begin{widetext}
\begin{equation}
	\begin{aligned}
		g^{(2)}_{{|\Phi \rangle}_{_{q}}}(0) = &\Bigg\{ {{{{|\cal N|}^2}}\, |_{_{q}}\langle \beta|\alpha\rangle_{_{q}}|^2} \Bigg( {|z|^4}+\frac{2g}{\sqrt{2}} \mathrm{\mbox{Re}} \bigg(  {\langle\hat{A}_{_{q}}\rangle_{_{w}}}   \left( \left(1+ q\right)  {z^*}  {|z|^2} + \left( {{q^2} {z^*}-z }\right) {|z|^4} \right) \bigg)\\ 
		&\hspace{-5mm}+\frac{g^2}{2}    {|{\langle\hat{A}_{_{q}}\rangle_{_{w}}}|^2} \bigg( \left( {q^2} + 2q + 1 - (1+q) {({z^2}+{z^*}^2)}\right) {|z|^2} - \left( q^2 (z^2 + {z^*}^2) - 2 q^3 - 2  q^{2} - q^{4}\right) {|z|^4} \\
		&\hspace{-5mm}+ \left( 1+q^5\right) {|z|^6}  \bigg)  \Bigg) \Bigg\} \times \Bigg\{ {{{{|\cal N|}^2}}\, |_{_{q}}\langle \beta|\alpha\rangle_{_{q}}|^2} \Bigg( {|z|^2}+\frac{2g}{\sqrt{2}} \mathrm{\mbox{Re}} \bigg( {\langle\hat{A}_{_{q}}\rangle_{_{w}}} \left( z^* + q {z^*} {|z|^2} - z {|z|^2}\right) \bigg) \\
		&\hspace{-5mm}+ \frac{{g^2}}{2} |{\langle\hat{A}_{_{q}}\rangle_{_{w}}} |^2 \bigg( 1-z^2 - {z^*}^2 - \left( q ( z^2 + {z^*}^2) - 2 q - q^2 \right) {|z|^2} + \left( 1+q^3\right)  {|z|^4} \bigg) \Bigg) \Bigg\} ^{-2}.
	\end{aligned}\label{eq:g2}
\end{equation}
\end{widetext}

\vfill\pagebreak
\onecolumngrid
\onecolumngrid
\begin{figure}[t!]
	\begin{center}
		\subfloat[]{\includegraphics[width=0.45\linewidth]{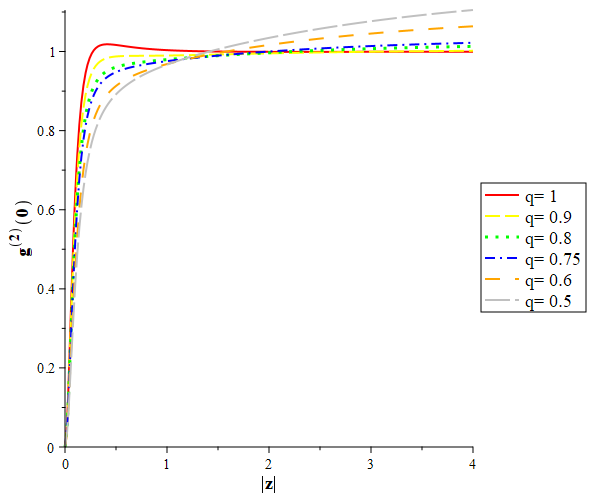}}\qquad
		\subfloat[]{\includegraphics[width=0.45\linewidth]{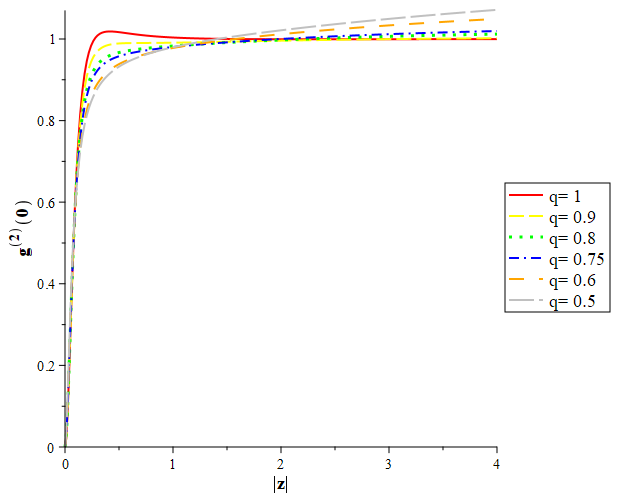}}\\
		\subfloat[]{\includegraphics[width=0.45\linewidth]{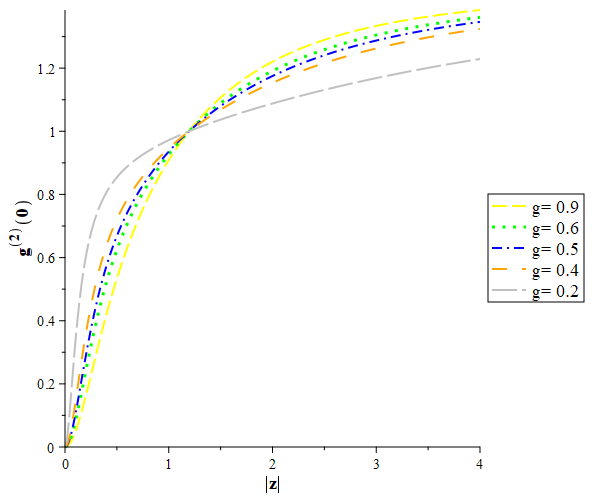}}\qquad
		\subfloat[]{\includegraphics[width=0.45\linewidth]{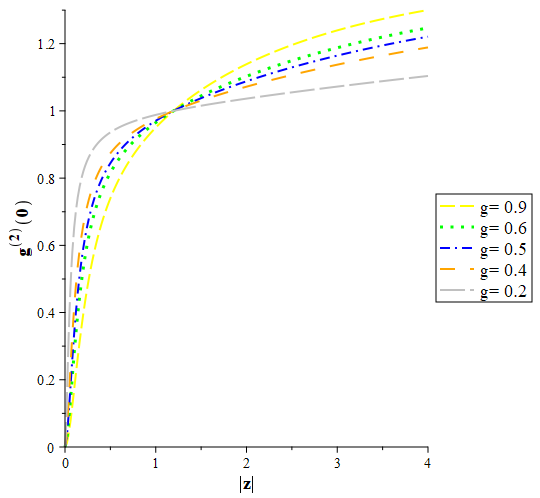}}
	\end{center}
	\caption{Second order $q$-correlation function of the $q$-post-selected probes with (a) the weak value $\langle\hat{X}_{_{1,q}}\rangle_{_{w}}$ and (b) $\langle\hat{X}_{_{2,q}}\rangle_{_{w}}$ both with $g=0.2$, $|\alpha|=2$, $|\beta|=0.5$ and $\theta=\frac{\pi}{2}$, $\lambda=\frac{\pi}{8}$, $\varphi=\frac{7\pi}{8}$. (c) for $\langle\hat{X}_{_{1,q}}\rangle_{_{w}}$ and (d) for $\langle\hat{X}_{_{2,q}}\rangle_{_{w}}$ both with $q=0.3$, $0<g<1$, $|\alpha|=2$, $|\beta|=0.5$ and $\theta=\frac{\pi}{2}$, $\lambda=\frac{\pi}{8}$, $\varphi=\frac{7\pi}{8}$.}\label{fig:Fig4}
\end{figure}
\twocolumngrid
\vfill\pagebreak
Substituting the two weak values $\langle\hat{A}_{_{q}}\rangle_{_{w}}=\langle\hat{X}_{_{1,q}}\rangle_{_{w}}$ and  $\langle\hat{A}_{_{q}}\rangle_{_{w}}=\langle\hat{X}_{_{2,q}}\rangle_{_{w}}$ in Eq.~\ref{eq:g2}, we obtain Figs.~\ref{fig:Fig4}(a)-(d), where classical photon bunching and pure quantum antibunching effects will be distinguished by $g^{(2)}_{{|\Phi \rangle}_{_{q}}}(0)>1$ and $g^{(2)}_{{|\Phi \rangle}_{_{q}}}(0)<1$, respectively~\cite{teich1988photon}.  
Fig.~\ref{fig:Fig4} shows that both bunching and antibunching effects are exhibited by the $q$-post-selected pointer states with a wide range of values of $q$ and $g$.

It is evident from Figs.~\ref{fig:Fig4} (a) and (b) that when the interaction parameter is fixed at $g=0.2$, decreasing the value of the deformation parameter $q$ results in a stronger antibunching effect in a small range of $|z|$. 
However, for the remaining values of $|z|$, the bunching effect will be the dominant mechanism. 

Fig.~\ref{fig:Fig4}(c) and (d) show that for $q=0.3$, increasing the value of $g$, stronger antibunching effect is exhibited by the pointer state for a limited values of the coherency parameter. For the remaining values of $|z|$, the bunching effect prevails, as seen in the case with interaction constant $g=0.2$.

Observing photon antibunching indicates that the distribution of photon numbers is squeezed in both cases, whether the deformation parameter or interaction coefficient is fixed~\cite{walls1983squeezed,teich1990squeezed}. 
However, when the deformation parameter is fixed and the interaction constant enhances, the photon number distribution is even more strongly affected.

\subsection{Quadrature Squeezing}
In this subsection, we consider the squeezing and expansion of position $\hat{X}=\frac{1}{\sqrt{2}} ({{\hat{{a}}}^\dagger}_{q}+\hat{a_q})$ and momentum $\hat{P}=\frac{i}{\sqrt{2}} ({{\hat{{a}}}^\dagger}_{q}-\hat{a_q})$ operators of the pointer state using the generalized Weyl-Heisenberg uncertainty inequality $\sigma_{XX} \sigma_{PP} \geqslant \frac{1}{4} |\langle[{\hat{X}}_q,{\hat{P}}_q] \rangle|^2$~\cite{buvzek1992superpositions}. 

To compute $\sigma_{XX}=\langle {{\hat{X}}^2}\rangle -{\langle {{\hat{X}}}\rangle}^2$ and $\sigma_{PP}=\langle {{\hat{P}}^2}\rangle -{\langle {{\hat{P}}}\rangle}^2$, we obtain the following expectation values in $q$-coherent state pointer 

\begin{widetext}
\begin{equation}\label{Xhat1}
	\begin{split}
	{_{q}\langle \Phi|}\hat{X}_{q}^{2} (\hat{P}_{q}^{2})|\Phi\rangle_{q} & = \frac{|\mathcal{N}|^{2}|_{q}\langle \beta|\alpha\rangle_{q}|^{2}}{2} \bigg\{ 1\pm z^{2} \pm {z^{*}}^{2} +(1+q) |z|^{2}
	 \pm \sqrt{2} g \mbox{Re} \Bigg( {\langle \hat{A}_{q}\rangle_{w}}\bigg( z^{*}-z \pm {z^{*}}^{3} \mp z^{3} +(1+q) (z^{*}\pm z)  \\
	& +(\pm q^{2} z +q(1+q)z^{*}-(1+q)z \mp z^{*}) |z|^{2}\bigg)\Bigg) \\
	&-\frac{g^{2}}{2}|\langle \hat{A}_{q}\rangle_{w}|^{2} \Bigg( \pm z^{4} \pm {z^{*}}^{4} +\bigg( (2 \mp 1)+(1\mp 1)q \mp q^{2} \bigg) (z^{2} +{z^{*}}^{2}) \\
	&+\bigg( \mp q^{3} +q^{2} +q \mp 1 \bigg) (z^{2}+ {z^{*}}^{2}) |z|^{2} +\bigg( -q^{3} -3 q^{2} -(3\mp 2)q \mp (-2\pm 1)\bigg) |z|^{2}  \\
	&  -\bigg(q^{4} +q^{3} \mp 2 q^{2} +q+1 \bigg) |z|^{4} -q-2  \bigg) \Bigg) \bigg\},
	\end{split}
\end{equation}

\begin{align}\label{Xhat2}
	{_{q}\langle \Phi|}\hat{X}_{_{q}} (\hat{P}_{_{q}}){{|\Phi \rangle}_{q}} &= \frac{{{|\cal N|}^{2} \, \,|_{_{q}}\langle \beta|\alpha\rangle_{_{q}}|^2} \, {i^{(\frac{1}{2}\mp \frac{1}{2})}}} {\sqrt{2}} \Bigg\{ z^*\pm z 
	\pm {\sqrt{2} \, g\, {i^{(\frac{1}{2} \mp \frac{1}{2})}}} \mbox{Re} (\mbox{Im})\bigg( {\langle \hat{A}_{_{q}}\rangle_{_{w}}} (1\pm{z^*}^2 -z^2 \nonumber\\
	&\quad +(q\mp 1) |z|^2 \bigg) -\frac{g^2}{2} {|\langle \hat{A}_{_{q}}\rangle_{_{w}}}|^2 \bigg({z^*}^3  \pm z^3 -q(z^* \pm z)+ (1\mp 1) (z-z^*) \nonumber\\
	&\quad +(\mp q^2 +q \mp 1)z {|z|^2} -(q^2 \mp q +1) {z^*} |z|^2 \bigg) \Bigg\}, 
\end{align}
and
\begin{align}
	\langle[{\hat{X}}_q,{\hat{P}}_q]\rangle &= {{|\cal N|}^{2} \, \,|_{_{q}}\langle \beta|\alpha\rangle_{_{q}}|^2}\Bigg\{ i (1-|z|^2 + q |z|^2) \nonumber\\
	&+i \frac{g}{\sqrt{2}} \mbox{Re} \Bigg( {\langle \hat{A}_{_{q}}\rangle_{_{w}}}  \bigg( z^3 +{z^*}^3 +q {z^*} -(2+q)z +(q^2 -q-1) {z^*} {|z|^2} -(q^2 +q -1) z {|z|^2}  \bigg)  \Bigg) \nonumber\\
	& -i \frac{g}{\sqrt{2}}  \mbox{Re} \Bigg(  {\langle \hat{A}_{_{q}}\rangle_{_{w}}}  \bigg( z^3 +{z^*}^3 -q( z + z^*) -({q^2} -q+1) z |z|^2 -(q^2-q+1) z^* |z|^2 \bigg)\Bigg) \nonumber\\
	&-i \frac{g^2}{2} \, |{\langle \hat{A}_{_{q}}\rangle_{_{w}}}|^2 \mbox{Re} \Bigg( \bigg( {z^*}^4 -z^4 -(q^3 +q^2 -q +1) |z|^2 
	+(q^2 +2q+1) z^2 -(1+q^2) {z^*}^2\nonumber\\
	& -(q^3 -q^2+q+1){{z^*}^2}{|z|^2}+(q^3 +q^2 -q+1){z^2}{|z|^2 -(q^4 -q^3+q-1){|z|^4}-q}\bigg) \Bigg) \Bigg\}.
\end{align}
\end{widetext}

\begin{figure}[t!]
	\begin{center}
		\subfloat[]{\includegraphics[width=0.36\linewidth]{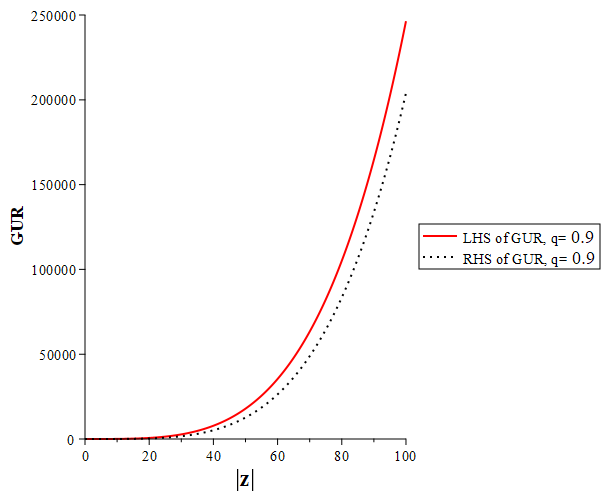}}\qquad
		\subfloat[]{\includegraphics[width=0.36\linewidth]{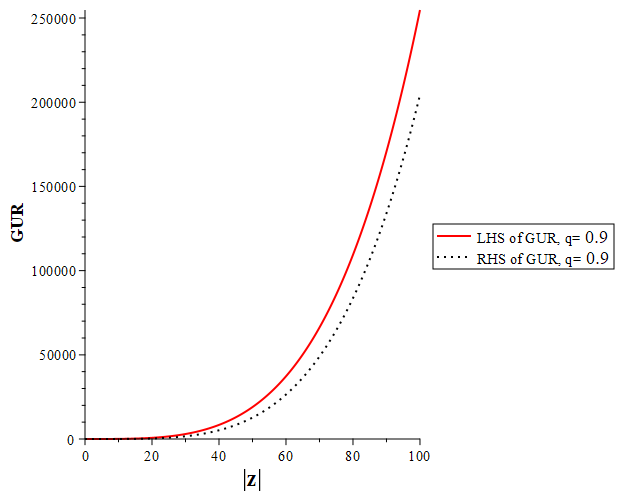}}\\ 
		\subfloat[]{\includegraphics[width=0.36\linewidth]{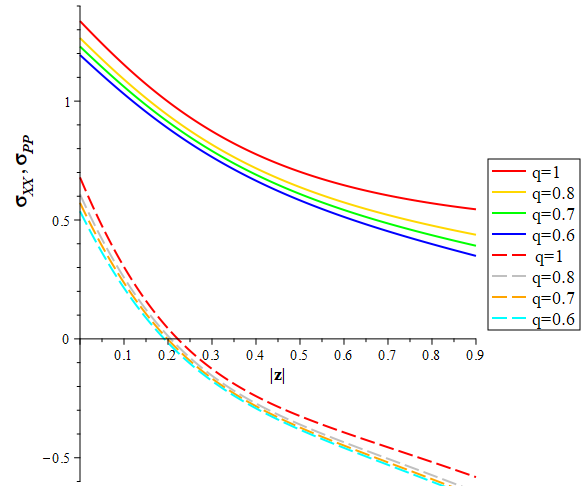}}\qquad
		\subfloat[]{\includegraphics[width=0.36\linewidth]{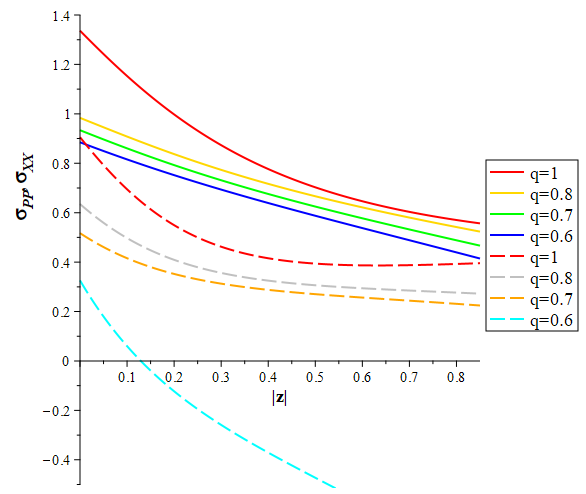}}\\
		\subfloat[]{\includegraphics[width=0.36\linewidth]{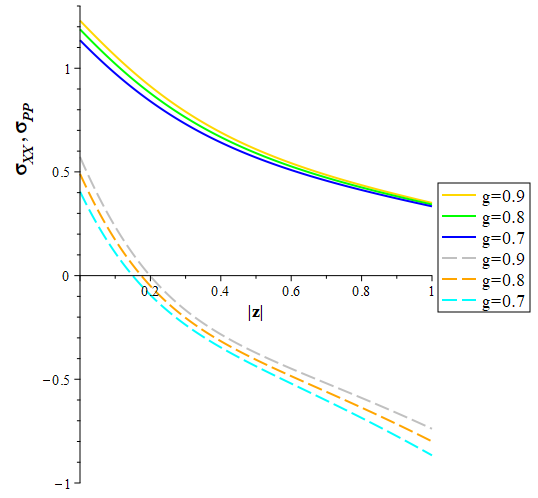}}\qquad
		\subfloat[]{\includegraphics[width=0.36\linewidth]{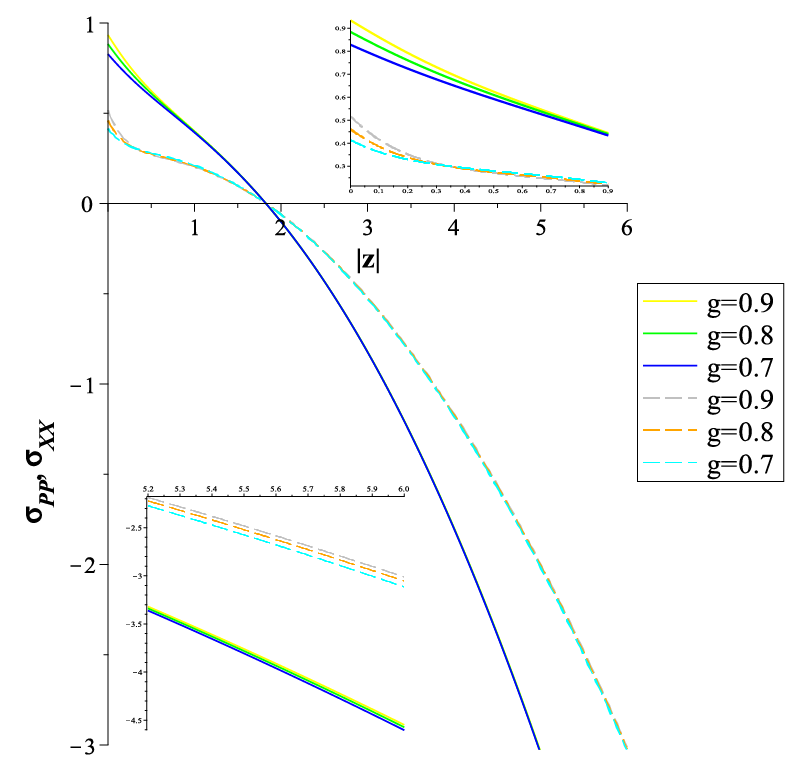}}
	\end{center}
	\caption{Plots of (a) generalized uncertainty relation (GUR) with coherence parameters $|\alpha|=4$, $|\beta|=2$, $\theta=\frac{\pi}{2}$, $\lambda=\frac{\pi}{8}$, $\varphi=\frac{7\pi}{8}$ for the observable $\hat{X}_{1}$ and (b) for the observable $\hat{X}_{2}$. (c) and (d) belong to the variances of the $q$-generalized position and momentum operators for various values of $0<q<1$ and $g=0.8$ with $|\alpha|=4$, $|\beta|=0.5$ and  $\theta=\frac{\pi}{2}$, $\lambda=\frac{\pi}{3}$, $\varphi=\frac{2\pi}{3}$. The solid red, yellow, green and blue curves show the variance in $X$, while the dashed red, gray, orange and cyan curves show the variance in $P$. (e) and (f) are related to the variances of the $q$-generalized position and momentum operators for various values of $0<g<1$ and $q=0.7$ with  $\theta=\frac{\pi}{2}$, $\lambda=\frac{\pi}{3}$, $\varphi=\frac{2\pi}{3}$.}\label{fig:Fig5}
\end{figure}

It should be noted that in Eqs.~\ref{Xhat1} and ~\ref{Xhat2}, the upper and lower signs are for $\hat{X}$ and $\hat{P}$, respectively.
Consequently, substituting the above values in the Weyl-Heisenberg uncertainty inequality, plots in Fig.~\ref{fig:Fig5} are obtained.
Figs.~\ref{fig:Fig5}(a) and (b) represent the $q$-generalized uncertainty relation for the $q$-coherent states of the meter with the weak values of the observables $\hat{A_1}=\hat{{X_1}_q}$ and $\hat{A_2}=\hat{{X_2}_q}$, respectively. It is clear that in both cases the Weyl-Heisenberg inequality is completely satisfied. 

\vfill\pagebreak
\onecolumngrid
\onecolumngrid
\twocolumngrid
\ \\
\vfill\pagebreak

As shown in Figs.~\ref{fig:Fig5}(c) and (d), the $\hat{X}$ quadrature is expanded and $\hat{P}$ is squeezed for the fixed value of the interaction parameter $g=0.9$ and various values of the deformation parameter. Furthermore, as the value of $q$ decreases, the degree of squeezing (expanding) in the $\hat{P}$ ($\hat{X}$) quadrature gets stronger (weaker).

Figs.~\ref{fig:Fig5}(e) and (f) clarify that in the case that the interaction parameter changes, quadrature squeezing is observed again in quadrature $\hat{P}$ and its conjugate quadrature, i.e. $\hat{X}$ is expanded. 
Moreover, the degree of squeezing gets stronger by decreasing the value of $g$, while the deformation parameter is fixed at $q=0.8$. 
These findings confirm that, unlike the ordinary (non-deformed) pointer coherent states, the uncertainties in the two quadratures are not equal. 

Hence, a compelling aspect of $q$-post-selected systems is that the level of squeezing can be adjusted through two parameters: $q$ and $g$. This parallels the idea that such systems possess two distinct supplementary degrees of freedom, namely the deformation parameter $q$ and the interaction coefficient $g$. This characteristic significantly contributes to the crucial role played by these states in quantum metrology.

The ideal squeezed states or intelligent states are special states that show squeezing in one quadrature and satisfy the uncertainty relation~\cite{gerry2023introductory}. 
Here, squeezing in one quadrature for various values of $q$ and $g$ is observed and the minimum uncertainty relation is also satisfied. 
Consequently, it can be inferred that the $q$-post-selected probe states also qualify as intelligent states.

\section{Concluding remarks}\label{sec:Conclusion}
We have leveraged a particular kind of post-selected weak measurement in which the initial and final states of the system, and the state of the pointer are replaced by the coherent states of a special type of harmonic oscillator, the $q$-deformed Arik-Coon harmonic oscillator with $0<q<1$. 
To the best of our knowledge, $q$-deformed systems have not been studied under the weak measurement regime.

Our efforts aimed to modify the weak measurement technique by measuring the $q$-generalized observables with a $q$-deformed coherent state meter, which best fits the non-ideal $\text{He-Ne}$ lasers.
We have shown how the strategic choices of the $q$-deformed initial and final post-selected states can result in large flexible weak values and optimized probes that are essential in quantum metrology for obtaining more precise measurements. 
$q$-dependency of the weak value makes it more tunable and flexible while setting a limit for its amplification value. 
Furthermore, the fidelity between the pre- and post-selected states was found to be dependent on $q$, varying with changes in the deformation parameter.
Plots of the probability of finding $n$ photons in a $q$-coherent state of the probe related to each weak value revealed that the height and position of the peaks for different values of $q$ deviates from the non-deformed case. 
$q$ and $g$-dependent antibunching effect and sub-Poissonian statistics are the additional nonclassical features exhibited simultaneously by the meter states, which cannot be observed with the non-deformed states. 
Number (amplitude) squeezing is another quantum characteristic of the $q$-post-selected pointer states as well. 

It was observed that the generalized Weyl-Heisenberg uncertainty relation is fully satisfied by the $q$-post-selected coherent pointer states.
Quadrature squeezing, the other quantum phenomena, is shown by the meter states. 
It was found that with the weak values of both types of $q$-deformed observables, both the $P$ and $X$ quadratures were squeezed with $g=cte$ and $q=cte$, respectively. 
Existence of both the quadrature and amplitude squeezing in this weak measurement regime is a special statistical trait shown by the $q$-post-selected pointers. 

The principal finding of this research indicates that through $q$-deformation of the post-selected weak measurement, enhanced nonclassicality of the probe state is attained, a pivotal factor in enhancing precision in quantum metrology.


\acknowledgements
This research is supported by T\"urkiye Scholarships, the Turkish Government Scholarship Program, under the Scholarship No. 23IR009840. F.O. acknowledges Personal Research Fund of Tokyo International University.

\bibliography{OQuL_v20240614}

\end{document}